\documentclass[%
 reprint,
 amsmath,amssymb,
 aps, showkeys
]{revtex4-2}

\usepackage{graphicx}
\usepackage{dcolumn}
\usepackage{bm}

\begin{document}

\title{Space-time-matter gravity as the origin of rotation in 4-D stationary and axisymmetric vacuum solutions}

\author{J.L. Hern\'andez--Pastora}
 \altaffiliation[Also at ]{https://ror.org/02f40zc51. \\ ORCID:orcig.org/0000-0002-3958-6083.}
 \email{jlhp@usal.es.}
\affiliation{Universidad de Salamanca,  Departamento de Matem\'atica Aplicada. Facultad de Ciencias. Instituto Universitario de F\'\i sica Fundamental y Matem\'aticas.
		Spain\\
}%

\date{\today}

\begin{abstract}

 Einstein's standard theory of General Relativity (GR) currently provides the most reliable description of all gravitational events in Astrophysics and Cosmology.
However, current Astronomy allows  measurements that contradict the predictions of GR in some gravitational scenarios  such as the accelerated expansion of the Universe,
the fast rotation of cluster of galaxies of the so strongly deflection of light  by  gravitational lenses. This has led the scientific community to propose modifications of the theory, and in particular to introduce the existence of dark matter and dark energy.

One of these modified theories of gravitation proposes, from  the same scheme that the geometrical theory of GR, a so called space-time-matter (STM) theory within a manifold of five dimensions where the resulting  metric were able  to measure adequately and adjusted to the current observations the most massive gravitational events.

Here we show that the gravitational features of a 4-D metric in the standard space-time can be understood as an induced  effect provided by a geometrical generalization of GR  into a 5-D manifold. In particular the rotation corresponding to an stationary and axisymmetric 4-D metric is explained as a manifestation of the existence of a new additional dimension of the manifold which can be related with the rest mass.
Furthermore the equivalence between a description of gravity in $4$ dimensions and the static vacuum solution constructed in a 5-D manifold could allow us to obtain stationary and axially symmetric solutions not known so far. These solutions can provide metrics that lead to determine those small observational corrections that are currently related to the existence of dark matter, but that in reality could be gravitational effects well explained within a geometrical theory generalizing Einstein's gravity.

\end{abstract}

\maketitle


\section{Introduction}

In \cite{Pastora} the field equations for  static and axially symmetric solutions in the framework of the non-compactified Kaluza-Klein \cite{KK} gravitational theory in a 5-dimensional manifold are explicitly showed.

In the present work we establish an equivalence between those field equations of this theory in 5 dimensions  (the STM theory) \cite{stm1},  and the corresponding ones into a 4 dimensional manifold with the  standard GR theory. In this way it is possible to identify the metric functions of each line element in both manifolds which verify the respective equations. Hence a correspondence is  established between the respective solutions of each gravitational theory. Moreover, these metrics, different from each other, would represent the same physical  gravitational scenario in both manifolds but with an eventually  different  description of the  physics and the gravitational effects on test particles in each manifold.

In the development of this relationship, a function $\Lambda$ is postulated which acts as a generator of two different scenarios: On the one hand, if $\Lambda=0$ then the identification of the field equations implies that $\omega=0$ (the rotation in the 4-D metric) and $\phi=cte$, i.e. the 4-D metric is static and the 5-D metric reduces to it (by necessarily making constant the potential related to the extra dimension $\phi$). In this case the family of static solutions of the standard GR is recovered.
Conversely, if $\Lambda \neq 0$ the 5-dimensional static solution is equivalent to a 4-dimensional stationary and axysimmetric solution (with rotation).

Within the non-compactified Kaluza-Klein approach, this STM theory relates the fifth new coordiante $x^4$ to the rest mass $m$ by means of a rescaling  $x^4=Gm/c^2$. Regardless of the interpretation made of the new coordiante, the relevance of this work consists on showing that the new dimension can be considered as a mechanism to explain  the rotation of an stationary 4-D metric as an effect derived from it in the same way that Kaluza-Klein theory \cite{KK} generates {\it matter} as an effect of the extra dimension.

\section{The identification of the field equations}

Let ${\mathbb{V}}_4$ be a Riemmanian manifold endowed with a stationary and axisymmetric
metric $g_{\alpha \beta}$. Let $\xi$ and $\eta$ the corresponding Killing
vectors associated with the time and axial isometries of this space-time 
respectively.
As is known, \cite{papa}, \cite{carter}, there exist coordinates
$\{x^{\alpha}\}=\{t,\rho,z,\phi\}$ ($\alpha=0..3$) adapted to both Killing vectors
and
the more general stationary and axially symmetric line element can be written as
follows:
\begin{widetext}
\begin{equation}
ds^2=-e^{2\sigma}\left(dt^2-\omega d\phi\right)^2+e^{2\Gamma
	-2\sigma}\left(dz^2+d \rho^2\right
)+J^2 e^{-2\sigma}d\phi^2 \ ,\label{metric}
\end{equation}
\end{widetext}
where $\sigma$, $\Gamma$, $\omega$ and $J$ are metric functions depending on
$x^1\equiv \rho$ and $x^2\equiv z$.
The Einstein's vacuum field equations for this metric
(\ref{metric}) are the following:

\begin{eqnarray}
&&\sigma ^{\prime \prime}+\ddot \sigma+\frac{\dot J}{J}\dot
\sigma+\frac{J^\prime}{J}\sigma^\prime+\frac{1}{2}\frac{e^{4\sigma}}{J^2}\left
((\omega^\prime)^2+\dot \omega ^2\right )=0 \label{4-D1} \\
&&\omega ^{\prime \prime}+\ddot \omega+4\omega^\prime\sigma^\prime+4\dot \omega
\dot
\sigma-\frac{J^\prime}{J}\omega ^\prime-\frac{\dot J}{J}\dot\omega=0\label{4-D2} \\
&&(\sigma ^\prime)^2+\dot \sigma^2+\frac{1}{4}\frac{e^{4\sigma}}{J^2}\left (\dot
\omega^2+(\omega^\prime)^2\right )+\Gamma ^{\prime \prime}+\ddot \Gamma=0
\label{4-D3} \\
&&2\dot \sigma \sigma ^\prime+\frac{\dot
	J^\prime}{J}-\frac{1}{2}\frac{e^{4\sigma}}{J^2}\dot \omega
\omega^\prime=\frac{J^\prime}{J}\dot \Gamma+\frac{\dot J}{J}\Gamma^\prime\label{4-D4} \\
&&(\sigma ^\prime)^2-\dot \sigma
^2+\frac{J^{\prime\prime}}{J}+\frac{1}{4}\frac{e^{4\sigma}}{J^2}\left(\dot
\omega^2-(\omega ^\prime)^2\right)=\frac{J^\prime}{J}\Gamma^\prime-\frac{\dot
	J}{J}\dot \Gamma \label{4-D5} \\
&& \ddot J+J^{\prime\prime}=0 \ ,
\label{4-D6}
\end{eqnarray}
where  $\prime$ and  $\dot{}$ denote  derivatives with respect to the coordinate
$\rho$ and $z$ respectively. The metric functions $\sigma$, $\omega$ and $J$ can
be characterized intrinsically from the Killing vectors by the following
scalars:
\begin{eqnarray}
\xi^{\alpha}\xi_{\alpha}&=&e^{-2\sigma} \quad , \quad
\xi^{\alpha}\eta_{\alpha}=e^{2\sigma}\omega \nonumber\\
\eta^{\alpha}\eta_{\alpha}&=&e^{-2\sigma}J^2-e^{2\sigma}\omega^2 \quad , \quad
2\xi_{[\alpha}\eta_{\nu]}\xi^{\alpha}\eta^{\nu}=J^2 \ .
\label{scalars}
\end{eqnarray}

An alternative theory of gravitation is developed in a huge number of research works \cite{stm1}, which is no more than a geometrical extension of Einstein gravity to a $5$-dimensional  also Riemmanian manifold ${\mathbb{V}}_5$ where an extra dimension is added by means of a new coordinate related with the rest mass. The STM theory developes the Kaluza mechanism and it is a
minimal extension of GR in the sense that there is no modification to the mathemathical structure of Einstein's theory just only changing the tensors
indices running over one more value.

We assume that there is no five-dimensional energy-momentum tensor and hence the Einstein equations in five dimensions  are (latin indices  run over 0-4) 
\begin{equation}
\hat S_{AB}=0=\hat R_{AB}
\label{einstein5}
\end{equation}
where $\hat S_{AB}$ and $\hat R_{AB}$ denote the Einstein and Ricci tensors respectively with respect to the five-dimensional metric. The absence of matter sources is explained  in the sense  that the universe in five dimensions is assumed to be empty and the idea to explain the matter in four-dimension is as a manifestation of pure geometry in higher ones. The five-dimensional Ricci tensor and Christoffel symbols are defined in terms of the metric as in four dimensions.

Let us note that the result obtained by Kaluza \cite{kaluza} to unify the electromagnetism and gravitation arises when  an electromagnetic $A_{\alpha}$ potential is coupled to the metric in the form $g_{\alpha \beta}+\phi^2 A_{\alpha}A_{\beta}$, and $g_{\alpha 4}=\phi^2 A_{\alpha}$.  In that case, if the scalar field $\phi$ is constant then the field equations provide directly the Einstein and Maxwell equations. Now we will work within the Brans-Dicke  theory \cite{bransdicke}, and we assume that there is no electromagnetic potential and consider the following 5-dimensional metric (latin indices  run over 0-4, and greek ones from $0$ to $3$)
\begin{equation}
d\hat s^2=\hat g_{AB}dx^A dx^B=g_{\alpha\beta}dx^{\alpha} dx^{\beta}+ \epsilon \phi^2 d\mu^2 \ ,
\label{metrica}
\end{equation}
where the  static and axisymmetric line element of the 4-dimensional metric in a general system of  cylindric coordinates is written as follows:
\begin{equation}
ds^2=-e^{2\sigma} dt^2+e^{2 b} d\rho^2+e^{2 c} dz^2+e^{2 d}  d\varphi^2,
\label{ERglobal}
\end{equation}
and  the metric functions  in these standard coordinates depend on $\rho$ and $z$, whereas the metric function $\phi$ is allowed to depend also on $t$ and $\varphi$ coordinates. The so-called {\it cylinder condition} is assumed with respect to the independence of all the metric functions on the new coordinate $x^4\equiv \mu$. The factor $\epsilon=\pm 1$ allows us to consider a timelike or spacelike signature for the fifth dimension .

A relevant result in differential geometry known as Campbell's theorem \cite{campbell} proves that Einstein’s field equations in 4-D (with an energy-momentum tensor) can always be smoothly (if locally) embedded in the 5-D Ricci equations (vacuum field equations in five dimensions). The four-dimensional Einstein equations $S_{\alpha \beta}=8\pi G T_{\alpha \beta}$ are automatically contained in the above vacuum equations (\ref{einstein5}) if the induced {\it matter} described by $T_{\alpha \beta}$ is understood as a manifestation of pure geometry in the higher-dimensional world  (which has been called the induced matter interpretation of Kaluza-Klein theory) by the following expression:
\begin{equation}
8\pi G T_{\alpha \beta}=\frac{1}{\phi} \nabla_{\beta}(\partial_{\alpha} \phi) \ ,
\label{eqcampo}
\end{equation}
and the new metric function $\phi$ satisfies a Klein-Gordon equation for a massless scalar field ($\Box\equiv g^{\alpha\beta}\nabla_{\beta}\partial_{\alpha}$ denotes  the standard D'Alambertian operator):
\begin{equation}
\Box \phi =0 \ ,
\label{dalambertiano}
\end{equation}
equation which is equivalent to the condition of null Ricci scalar since the trace of the energy-momentum tensor $T_{\alpha \beta}$ vanishes.

The corresponding field equations from the STM theory  are the following ones  \cite{Pastora}:
\begin{widetext}
\begin{eqnarray}
&&E_{\rho \rho}+E_{z z}=0 \ , \ \phi_{\rho \rho}+\phi_{z z}=0 \ , \ \nabla E \cdot \nabla \phi=0 \label{5-D1}\\
&&\sigma_{\rho \rho}+\sigma_{z z}=\nabla \sigma \cdot \frac{\nabla \phi}{\phi}-\nabla \sigma \cdot \frac{\nabla E}{E} \label{5-D2}\\
&&\gamma_{\rho \rho}+\gamma_{z z}=\nabla \sigma \cdot \frac{\nabla \phi}{\phi}-\left(\nabla \sigma \right)^2 \ , \label{5-D3}
\end{eqnarray}
where the notation ${\displaystyle \nabla \sigma \nabla \phi \equiv \sigma_{\rho}\phi_{\rho}+\sigma_z \phi_z}$, ${\displaystyle (\nabla \sigma)^2\equiv \sigma_{\rho}^2+\sigma_z^2}$ is used, and the metric function $\gamma$ should be
\begin{equation}
\gamma=\sigma+\frac 12 \ln\left((\nabla \phi)^2 \right)+ h (z)
\label{gamma5-D}
\end{equation}
and it does verify the system of equations

\begin{eqnarray}
&& \gamma_{z}\left(\frac{E_{\rho}}{E}-\frac{\phi_{\rho}}{\phi} \right)+\gamma_{\rho}\left(\frac{E_{z}}{E}-\frac{\phi_{z}}{\phi}\right)= 2\sigma_z\sigma_{\rho}-\frac{\phi_{z\rho}}{\phi}+\frac{E_{z\rho}}{E}-\sigma_z\frac{\phi_{\rho}}{\phi}-\sigma_{\rho}\frac{\phi_z}{\phi}\nonumber \\
& &  - \gamma_{z}\left(\frac{E_{z}}{E}-\frac{\phi_{z}}{\phi} \right)+\gamma_{\rho}\left(\frac{E_{\rho}}{E}-\frac{\phi_{\rho}}{\phi}\right)= \sigma_{\rho}^2-\sigma_{z}^2-\frac{\phi_{\rho\rho}}{\phi}-\frac{E_{zz}}{E}+\sigma_z\frac{\phi_{z}}{\phi}-\sigma_{\rho}\frac{\phi_{\rho}}{\phi}\nonumber \\
\label{cuadra5-D}
\end{eqnarray}
and the 5-dimensional metric  is
\begin{equation}
d\hat s^2=- e^{2 \sigma} dt^2+e^{2 h}\left( \nabla \phi\right)^2\left( d\rho^2+ dz^2\right)+E^2 e^{-2\sigma}  d\varphi^2+\epsilon \phi^2 d\mu^2,
\label{metrica5-D}
\end{equation}
\end{widetext}

If we identify the metric function $J$ with the corresponding $E$ in the 5-D metric, as well as both metric functions $\sigma$, and $\Gamma=\gamma$ then the field equations (\ref{4-D4}),(\ref{4-D5}) recover exactly the system of equations (\ref{cuadra5-D}) iff $\dot \omega=\omega^{\prime}=0$, i.e. the rotation vanishes and hence the 4-D metric becomes static. This fact occurs because  the quadrature (\ref{cuadra5-D}) with the function  $\Gamma$ (\ref{gamma5-D})  recovers exactly the quadrature in $4-$dimensional field equations (\ref{4-D4})-(\ref{4-D5}). In addition, $\omega=0$ implies that $\phi$ must be constant when we identifiy the field equations (\ref{5-D2})-(\ref{5-D3}) with the corresponding (\ref{4-D1}) and (\ref{4-D3}). Therefore the 5-dimensional metric becomes the standard metric in four dimensions and we recover the same axially symmetric static solution.

But we can manage with  a new function namely $\Lambda$ into the identification of both metric functions $\Gamma$ and $\gamma$ of the corresponding metrics: $\Gamma=\gamma+\Lambda$ which leads to
\begin{eqnarray}
\frac{\nabla \phi}{\phi}\nabla \sigma&=& -\frac 14 \frac{e^{4\sigma}}{E^2}(\nabla \omega)^2-\triangle\Lambda\label{identi1}\\
\frac{\nabla \phi}{\phi}\nabla \sigma&=& -\frac 12 \frac{e^{4\sigma}}{E^2}(\nabla\omega)^2\label{identi2}
\end{eqnarray}
if the equations (\ref{4-D1}) and (\ref{4-D3}) are compared with the corresponding field equations in the 5-D metric, (\ref{5-D2}) and (\ref{5-D3}), as well as the following conclusion
\begin{eqnarray}
-\frac 12 \frac{e^{4\sigma}}{E^2}\dot\omega \omega^{\prime}&=&\dot \Lambda \frac{\phi^{\prime}}{\phi}+ \Lambda^{\prime} \frac{\dot \phi}{\phi}\label{idenomegapro}\\
\frac 14 \frac{e^{4\sigma}}{E^2}(\dot\omega^2- \omega^{\prime 2})&=&-\dot \Lambda \frac{\dot\phi}{\phi}+ \Lambda^{\prime} \frac{\phi^{\prime}}{\phi}\label{idenomegadif}
\end{eqnarray}
when comparing the systems of equation (\ref{cuadra5-D}) and (\ref{4-D4})-(\ref{4-D5}) in both metric for the metric functions $\gamma$ and $\Gamma$ respectively.

Therefore we need to make compatible the equations (\ref{idenomegapro})-(\ref{idenomegadif}) with the following expresions derived from (\ref{identi1})-(\ref{identi2}):
\begin{eqnarray}
\triangle \Lambda&=& \frac 14 \frac{e^{4\sigma}}{E^2}(\nabla \omega)^2 \label{sum1}\\
\frac{\nabla \phi}{\phi}\nabla \sigma&=& -\frac 12 \frac{e^{4\sigma}}{E^2}(\nabla\omega)^2\label{sum}
\end{eqnarray}

Before starting with the resolution of those equations we proceed to make a change of function, and a comment with respect to the last  equation of the system of field equations in 4-D still no mentioned: let us redefine the metric function $\omega $ as follows:
\begin{equation}
\dot \omega=J e^{-4\sigma} W^{\prime}, \qquad \omega^{\prime}=-J  e^{-4\sigma} \dot W \label{W}
\end{equation}
in such a way that (as it is known) the equation (\ref{4-D2}) is just the integrability condition of this redefinition (\ref{W}). Hence the new metric function $W$ must verify:
\begin{equation}
\triangle W-4\nabla\sigma\nabla W+\nabla W\frac{\nabla J}{J}=0
\label{ci}
\end{equation}

Therefore, the set of equations we should solve, in addition to the above (\ref{ci}), is the following:
\begin{eqnarray}
(\nabla W)^2&=& -2 e^{4\sigma} \frac{\nabla \phi}{\phi}\nabla \sigma \label{EQ1}\\
\dot W \ W^{\prime}&=&2 e^{4\sigma}\left(\dot \Lambda \frac{\phi^{\prime}}{\phi}+ \Lambda^{\prime} \frac{\dot \phi}{\phi}\right)\label{EQ2}\\
\dot W^2- W^{\prime 2}&=&4 e^{4\sigma}\left(\dot \Lambda \frac{\dot\phi}{\phi}- \Lambda^{\prime} \frac{\phi^{\prime}}{\phi}\right)\label{EQ3}\\
\triangle \Lambda&=& \frac 14 e^{-4\sigma}(\nabla W)^2 \ . \label{EQ4}
\end{eqnarray}

\subsection{The metric function $W$}

From (\ref{EQ1})-(\ref{EQ3}) it is easy to see that 
\begin{eqnarray}
\dot W^2&=&2 e^{4\sigma}\left(-\frac 12 \frac{\nabla \phi}{\phi}\nabla \sigma+\dot \Lambda \frac{\phi^{\prime}}{\phi}- \Lambda^{\prime} \frac{\dot \phi}{\phi}\right)\label{Wdot}\\
W^{\prime 2}&=&2 e^{4\sigma}\left(-\frac 12 \frac{\nabla \phi}{\phi}\nabla \sigma-\dot \Lambda \frac{\phi^{\prime}}{\phi}+ \Lambda^{\prime} \frac{\dot \phi}{\phi}\right)\label{Wprime}
\end{eqnarray}
iff the function $\Lambda$ verifies, in adittion to (\ref{EQ4}), the following equation
\begin{equation}
(\nabla \Lambda)^2=\frac 14 \frac{\left( \nabla \phi\nabla \sigma\right)^2}{(\nabla \phi)^2} \ . \label{lambdacondnabla}
\end{equation}
Let us note that (see \cite{Pastora} for details) ${\displaystyle \frac{\nabla \phi}{\phi} \nabla \sigma=\beta \frac{(\nabla \phi)^2}{\phi^2}}$ once $\sigma$ is solved from the 5-D field equations,  with 
\begin{equation}
\beta=\frac{\phi^2}{\sqrt{(E^2+\phi^2\mp2k)^2\pm 8kE^2}} \ .
\end{equation} 

\vskip 5mm

On the first hand,  the integration of equations (\ref{Wdot}) and (\ref{Wprime}) for the metric function $W$ requires the integrability condition (\ref{ci}) which can be solved as follows: it is easy to see that this integrability condition can be written 
\begin{equation}
\dot W \dot X_{+}+W^{\prime} X_{-}^{\prime}=0
\label{wxy}
\end{equation}
where
\begin{equation}
X_{\pm}\equiv -2\sigma+\ln E+\ln \sqrt{G\pm D} \ ,
\end{equation}
$G$ and $D$ being the following functions defined from the expressions (\ref{Wdot}) and (\ref{Wprime})  $\dot W=e^{2\sigma} \sqrt{G+D}$, $W^{\prime}=e^{2\sigma} \sqrt{G-D}$:
\begin{equation}
G\equiv -\beta \frac{(\nabla \phi)^2}{\phi^2} , \qquad D\equiv 2 \left( \dot \Lambda \frac{\phi^{\prime}}{\phi}- \Lambda^{\prime} \frac{\dot \phi}{\phi}\right) \ .
\label{GyD}
\end{equation}

The  general solution of the equation (\ref{wxy}) for an arbitrary function $f(\rho,z)$ is:
\begin{eqnarray}
\dot W&=&\frac{e^{4\sigma}}{E}\left(k_1(\rho)+\int f(\rho,z) E e^{-2\sigma} dz \right)\label{eme}\\
W^{\prime}&=&\frac{e^{4\sigma}}{E}\left(k_2(z)-\int f(\rho,z) E e^{-2\sigma} d\rho \right)\label{ene}
\end{eqnarray}
Let us note that, in fact, the integrability condition (\ref{ci}) is just an anisotropic heat equation type and it is equivalent to 
\begin{equation}
0=\partial_{\rho}\left(W^{\prime} J e^{-4\sigma}\right)+\partial_{z}\left(\dot W J e^{-4\sigma}\right) \ ,
\end{equation}
and hence a function $K$ should exist verifying that ${\displaystyle \dot K=W^{\prime}J e^{-4\sigma}}$ and ${\displaystyle K^{\prime}=- \dot W J e^{-4\sigma}}$ to ensure integrability: this function $K$ is just the one whose derivatives appear in the brackets of above equations (\ref{eme})-(\ref{ene}) (with $J=E$):
\begin{eqnarray}
K^{\prime}&=&-k_1(\rho)-\int f(\rho,z) E e^{-2\sigma} dz \label{Kdot}\\
\dot K&=&k_2(z)-\int f(\rho,z) E e^{-2\sigma} d\rho\label{Kprime}
\end{eqnarray}
and hence the function $f(\rho,z)$ could be determined in terms of $\phi$ and $\Lambda$ from  (\ref{Wprime}):
\begin{equation}
f(\rho,z)=-\frac{e^{2\sigma}}{E}\frac{d}{d\rho}\left(Ee^{-2\sigma}\sqrt{G-D} \right) \ .
\end{equation}

\vskip 5mm

\subsection{The metric function $\Gamma$}

On the second hand we have to garantee the existence of solution for the pair of equations (\ref{EQ4}) and (\ref{lambdacondnabla}) with respect to the new function $\Lambda$. Those equations read  in terms of the 5-D metric function $\phi$ as follows:
\begin{equation}
\triangle \Lambda=-\frac 12 \beta \frac{(\nabla \phi)^2}{\phi^2}  , \qquad (\nabla \Lambda)^2=\frac 14 \beta^2 \frac{(\nabla \phi)^2}{\phi^2} \ . \label{eqdeG}
\end{equation}
Hence we need to solve the system of equations
\begin{eqnarray}
x^2+y^2&=&-\frac{\beta}{4}G \label{grad}\\
x^{\prime}+\dot y&=&\frac G2 \label{lapla}\\
\dot x-y^{\prime}&=&0 \label{cint}
\end{eqnarray}
where the notation $x\equiv \Lambda^{\prime}$ and $y\equiv\dot \Lambda$ has been used.

The condition for the gradient of $\Lambda$ (\ref{grad}) is solved in this way:
\begin{equation}
x={\mathbb{A}} \cos \nu, \qquad y={\mathbb{A}} \sin \nu\label{xyplanas}
\end{equation}
with ${\mathbb{A}}\equiv \sqrt{-\frac{\beta}{4}G}$ and $\nu$ being some  function  that is determined with the others two conditions as follows: the integrability condition (\ref{cint}) and the condition on the Laplacian of $\Lambda$ (\ref{lapla}) requiere respectively:
\begin{eqnarray}
a_{-} \cos \nu-a_+ \sin \nu&=&0\\
a_{+} \cos \nu+a_- \sin \nu&=&\frac G2 \ ,
\end{eqnarray} 
where $a_-\equiv \dot {\mathbb{A}}- {\mathbb{A}} \nu^{\prime}$ and $a_+ \equiv {\mathbb{A}}^{\prime}+ {\mathbb{A}} \dot \nu$. The unique solution is given by
\begin{equation}
a_+=\frac G2 \cos \nu, \qquad a_-=\frac G2  \sin \nu \ ,
\end{equation}
which means
\begin{equation}
a_-^{2}+a_+^{2}=\left(\frac G2 \right)^2 \ ,
\end{equation}
or equivalently, with $a\equiv \ln {\mathbb{A}}$
\begin{equation}
(\dot\nu+a^{\prime})^{2}+\left(\nu^{\prime}-\dot a\right)^{2}=\frac{(\nabla \phi)^2}{\phi^2} \ . \label{ultima}
\end{equation}
The solution of this equation (\ref{ultima}) is
\begin{equation}
\dot \nu=\frac{\phi^{\prime}}{\phi}-a^{\prime} , \qquad \nu^{\prime}=-\frac{\dot \phi}{\phi}+\dot a
\end{equation}
iff $\triangle a=-\frac{(\nabla \phi)^2}{\phi^2}$ which is the integrability condition for $\nu$. With a little bit of algebra it is easy to prove that this condition is verified since \cite{Pastora}
\begin{equation}
a\equiv\ln {\mathbb{A}}=\frac 12 \ln \left(\frac{\phi^2 (\nabla \phi)^2}{4 \left[(E^2+\phi^2\mp2k)^2\pm 8kE^2\right]} \right)
\end{equation}

\section{The theorems}

\subsection{Theorem 1}
{\it A relationship exists between the stationary and axisymmetric solutions of Einstein's vacuum-equations in a Riemmanian manifold ${\mathbb{V}}_4$ and the static and axisymmetric solutions of the  space-time-matter gravity based in a non-compactified Kaluza-Klein theory in a 5-D manifold.}

{\bf Proof:}
Let  $\{\sigma,\Gamma, J, \omega\}$ be the metric functions of a line element (\ref{metric}) which are solutions of the field equations (\ref{4-D1})-(\ref{4-D6}) derived from   standard GR theory  in a 4-D manifold, and let  $\{\sigma^{(5)}, \gamma, E, \phi\}$ be the respective  functions of a metric (\ref{metrica5-D}) defined in a 5-D manifold corresponding to a STM theory of gravitation with the field equations (\ref{5-D1})-(\ref{cuadra5-D}).  If we establish the following identification between both set of functions
\begin{equation}
\sigma=\sigma^{(5)} ,\qquad \Gamma=\gamma+\Lambda ,\qquad J=E \ ,
\end{equation}
then the corresponding field equations of both metrics are identical:
\begin{equation}
\begin{matrix}
  [ \qquad   4-D \qquad ]& & [ \qquad   5-D \qquad ]\\
\triangle J=0 & | & \triangle E=0\\
\triangle  \sigma:   Eq.(\ref{4-D1}) & | & \triangle \sigma^{(5)}: Eq.(\ref{5-D2})\\
\triangle  \Gamma:   Eq.(\ref{4-D3}) & | & \triangle \gamma: Eq.(\ref{5-D3})\\
\Gamma^{'s}: Eqs.(\ref{4-D4})-(\ref{4-D5}) & | & \gamma^{'s}: Eqs.(\ref{cuadra5-D})
\end{matrix} \label{espejo}
\end{equation} 
iff equations (\ref{idenomegapro}) and (\ref{idenomegadif}) are compatible with (\ref{sum1}), (\ref{sum}) or in other words iff the set of equations (\ref{EQ1})-(\ref{EQ4}) can be solved.

Each set of field equations (\ref{espejo}) is  completed  with the equation (\ref{4-D2}) for the metric function $\omega$ in 4-D and $\phi$ (\ref{5-D1}) from 5-D manifold respectively.

As it has been showed previously the solution of the system of equations (\ref{EQ1})-(\ref{EQ4}) is given by the functions $\omega$ (or equivalent $W$) and $\Gamma$ (or equivalently $\Lambda$). In section 2.1 is shown the integrability of $W$ leading to
\begin{equation}
\omega=\int E e^{-2\sigma} \sqrt{G-D} \ dz
\label{finalO}
\end{equation}
and the function $\Gamma$ is  obtained in section 2.2 by means of the expression
\begin{equation}
\Lambda=\int {\mathbb{A}} \cos \nu \ d\rho \ ,
\label{finalL}
\end{equation}
 where $G$ and $D$ are given in (\ref{GyD}), and  
 \begin{equation}
 \nu=\int\left( \frac{\phi^{\prime}}{\phi} -\frac{{\mathbb{A}}^{\prime}}{{\mathbb{A}}}\right) \ dz \ . \label{nu}
 \end{equation}
 
 \hfill  {$ \Box$}

\subsection{Theorem 2}

{\it The rotation of an stationary and axisymmetric solution of Einstein's field equations can be obtained as a manifestation of an extra coordinate in a ${\mathbb{V}}_5$ manifold where gravity is explained by means of an static and axisymmetric space-time-matter metric.}

{\bf Proof:} The function $\Gamma$ is a solution of (\ref{eqdeG}) given by the above equation (\ref{finalL}) which is directly related with the function $\phi$ of the extra dimension, as well as $W$ (or equivalently $\omega$) solved by means of $G$ and $D$ (\ref{GyD}).

Therefore  a relationship is established between the function $\omega$  asociated to the rotation in 4-D metric
 and the function $\phi$ connected with the fifth dimension, in such a way that two different scenarios are contemplated:

{\it i)} If $\Lambda=0$ then $\omega=0$ and  $\phi=cte$, and consequently both metrics are identical and they recover the static and axially symmetric space-time in GR. This is so because the field equations (\ref{4-D4}),(\ref{4-D5}) recover exactly the system of equations (\ref{cuadra5-D}) in that case, as well as the quadrature (\ref{cuadra5-D}) with the function  $\Gamma$ (\ref{gamma5-D})  recovers exactly the quadrature in 4-dimensional field equations (\ref{4-D4})-(\ref{4-D5}). In addition, $\omega=0$ implies that $\phi$ must be constant when we identifiy the field equations (\ref{5-D2})-(\ref{5-D3}) with the corresponding (\ref{4-D1}) and (\ref{4-D3}).

{\it ii)} Otherwise, if $\Lambda\neq0$ then  a non vanishing rotation $\omega$ arises at the 4-D metric. In this case $\omega\neq0$ is directly related with the new metric function $\phi$ by means of the equations (\ref{finalL})-(\ref{finalO}). Hence, we can hold that this physical feature (the existence of rotation) described by an stationary and axysimmetric solution of standard GR in $4$ dimensions can be also described by  an static metric in $5$ dimensions provided by  the STM theory of gravitation.

The extra dimension  generates in some sense the rotation of the stationary 4-dimensional metric, like a Kaluza-Klein geometrical mechanism of providing {\it matter}.

\hfill  {$ \Box$}

\section{Conclusions}

The most relevant results addressed and resolved in this paper  are as follows:

$\bullet $ A relationship  between axially symmetric  and stationary metrics in four dimensions and static and axially symmetric metrics in five dimensions is established.

$\bullet $ An alternative method  to obtain axially symmetric stationary metrics in GR is provided  from the solution of the field equations of the STM theory in five dimensions. Stationary solutions of axially symmetric Einstein's equations are obtained in GR by means of the Ernst's equation \cite{ernst1}. The equivalence presented here provides a mechanism to obtain new solutions. This is a goal to which much effort and time has been devoted in the field of exact solutions.

$\bullet $ It is postulated that the gravitational characteristics of an stationary and axially symmetric metric in four dimensions can be explained from an static 5-dimensional STM theory, so that the extra dimension can be understood as the generator of rotation, just as the Kaluza-Klein theory explains the induced matter effect from geometry.

\vskip 2mm

It is accepted that Einstein's GR provided a new paradigm  for explainnig the gravitation. And the assumption of a geometrical description of gravity leads to  predictions of corrections to the classical Newtonian Gravity by incorporating a new coordinate $ct$ to the spatial metric in  3-D.  The STM theory of gravity in 5-D has the advantage of being a geometrical generalisation of GR to one more dimension by resizing the rest mass like the variable time $t$  was done in Einstein's theory, an in addition there is no mathematical modification of the structure of Einstein's theory. But regardless of the interpretation of the physical meaning attributed to the new dimension,  this extended geometrical description of gravity  provides a mechanism for explaining gravitational effects that eventually do not appear in a 4-dimensional manifold and that allows us to obtain corrections to the gravitational description of GR. 

In this paper we show that the gravitational effects and physical characteristics of a metric in a 4-dimensional manifold can be described and understood as a consequence of an extended  gravitational theory in a 5-dimensional manifold. As has been already said current Astronomy allows  measurements that contradict the predictions of GR in some gravitational scenarios \cite{cosmo} such as the accelerated expansion of the Universe,
the fast rotation of cluster of galaxies of the so strongly deflection of light  by  gravitational lenses. This has led the scientific community to propose modifications of the theory, and in particular to introduce the existence of dark matter and dark energy. So a question that now arises may be whether it is really necessary to impose dark matter on the standard theory of gravity or whether  the observational  effects due to the apparent mass defect in the universe could be simply explained by a gravitational theory with one more dimension. Let me conclude with an analogy to illustrate the new insights into the Physiscs when a new dimension is outlined: Two dimensional
Newtonian physicists explaining the gravitational attraction at a curved region of space (which they do not
know about because they assure that the space is flat) would speak of a strange attracting force at that point, while
two-dimensional relativistic physicists (who does really know the curvature of the two-dimensional space-time but they
do not know about the third dimension) would speculate about the existence of a black hole (or whatever compact object) at the center of that region
of two-dimensional space that is curving it. The three-dimensional explanation of the phenomenon would provide an
effect induced by the geometry of three-dimensional space-time that does not necessarily imply the existence of a
stellar object: the two-dimensional space time is no longer empty and the two-dimensional Einstein equations can be
solved with a material content induced (but not physically real) by means of the extra dimension.

%
%

\begin{acknowledgments}
This  work  was  partially supported by the 
Grant PID2021-122938NB-I00 funded by MCIN/AEI/
10.13039/501100011033 and by “ERDF A way of making Europe”, as well as  the Consejer\'\i a
de Educaci\'on of the Junta de Castilla y Le\'on under the Research Project Grupo
de Excelencia GR234 Ref.:SA096P20 (Fondos Feder y
en l\'\i nea con objetivos RIS3).
\end{acknowledgments}


\end{document}